\documentclass[runningheads]{llncs}
\usepackage{graphicx}
\usepackage{todonotes}
\usepackage{subfigure}
\usepackage[misc]{ifsym}

\begin{document}
\title{Deep Learning for Automatic Spleen Length Measurement in Sickle Cell Disease Patients}
\author{Zhen Yuan\inst{1}$^{(\textrm{\Letter})}$ \and
Esther Puyol-Antón\inst{1} \and
Haran Jogeesvaran\inst{2} \and
Catriona Reid\inst{2} \and
Baba Inusa\inst{2} \and
Andrew P. King\inst{1}}
%
\authorrunning{Zhen Yuan et al.}
%
\institute{School of Biomedical Engineering and Imaging Sciences, King's College London, London, UK 
\\
\email{zhen.1.yuan@kcl.ac.uk} \and
Evelina Children’s Hospital, Guy’s and St Thomas’ NHS Foundation Trust, London, UK}
\titlerunning{Deep Learning for Automatic Spleen Length Measurement}%
\maketitle              
\begin{abstract}
Sickle Cell Disease (SCD) is one of the most common genetic diseases in the world. Splenomegaly (abnormal enlargement of the spleen) is frequent among children with SCD. If left untreated, splenomegaly can be life-threatening. The current workflow to measure spleen size includes palpation, possibly followed by manual length measurement in 2D ultrasound imaging. However, this manual measurement is dependent on operator expertise and is subject to intra- and inter-observer variability. We investigate the use of deep learning to perform automatic estimation of spleen length from ultrasound images. We investigate two types of approach, one segmentation-based and one based on direct length estimation, and compare the results against measurements made by human experts. Our best model (segmentation-based) achieved a  percentage length error of 7.42\%, which is approaching the level of  inter-observer variability (5.47\%-6.34\%). To the best of our knowledge, this is the first attempt to measure spleen size in a fully automated way from ultrasound images.

\keywords{Deep Learning  \and Sickle Cell Disease \and Spleen Ultrasound Images.}
\end{abstract}
\section{Introduction}
Sickle Cell Disease (SCD) is one of the most common genetic diseases in the world, and its prevalence is particularly high in some parts of the developing world such as sub-Saharan Africa, the Middle East and India \cite{chakravorty2015sickle,piel2013global,grosse2011sickle}. In the United States, 1 in 600 African-Americans has been diagnosed with SCD \cite{angastiniotis1998global,Biswasf4676}, and it was reported that 1 in every 2000 births had SCD during a newborn screening programme in the United Kingdom \cite{modell2007epidemiology,streetly2009implementation}. This multisystem disorder results in the sickling of red blood cells, which can cause a range of complications such as micro-vascular occlusion, haemolysis and progressive organ failure \cite{chakravorty2015sickle,piel2017sickle,inusa2016introductory}.

The spleen is an essential lymphatic organ of the human body, which plays the role of a filter that cleans blood and adapts immune responses against antigens and microorganisms. Splenomegaly (abnormal enlargement of the spleen) is frequent among children with SCD. Splenomegaly, if left untreated, can be a serious and life-threatening condition, and so SCD patients typically have their spleen size measured at routine clinical appointments \cite{brousse2014spleen}.

The typical workflow for measuring the size of the spleen includes palpation, possibly followed by manual length measurement from a 2D ultrasound examination. However, this workflow suffers from a number of drawbacks. First of all, palpation is relatively crude and non-quantitative, and only allows the clinician to make a preliminary judgement as to whether further ultrasound examination is required. Second, spleen measurement from ultrasound requires sonographer expertise and is subject to significant intra- and inter-observer variability \cite{rosenberg1991normal}. Furthermore, in parts of the developing world, where there is a high prevalence of SCD, there is often a shortage of experienced sonographers to perform this task. 

In recent years, deep learning models have been proposed for automatic interpretation of ultrasound images in a number of applications~\cite{ghorbani2020deep,namburete2017robust,kuo2019automation}. In~\cite{ghorbani2020deep}, the EchoNet model was proposed for segmentation of the heart's anatomy and quantification of cardiac function from echocardiography. In \cite{namburete2017robust}, a Convolutional Regression Network was used to directly estimate brain maturation from three-dimensional fetal neurosonography images. Recently, \cite{kuo2019automation} developed a model based on ResNet to estimate kidney function and the state of chronic kidney disease from ultrasound images. However, although automatic spleen segmentation has been attempted from magnetic resonance and computed tomography images  \cite{mihaylova2016brief}, automatic interpretation of ultrasound images of the spleen remains challenging due to the low intensity contrast between the spleen and its adjacent fat tissue. To the best of our knowledge, there is no prior work on automated spleen segmentation or quantification from ultrasound. In this paper, we propose the use of deep learning for automatic estimation of the length of the spleen from ultrasound images. We investigate two fully automatic approaches: one based upon image segmentation of the spleen followed by length measurement, and another based on direct regression against spleen length. We compare the results of these two methods against measurements made by human experts.

\section{Methods}
\begin{figure*}[!t]\center
	\subfigure[U-Net based] {\includegraphics[width=1\textwidth]{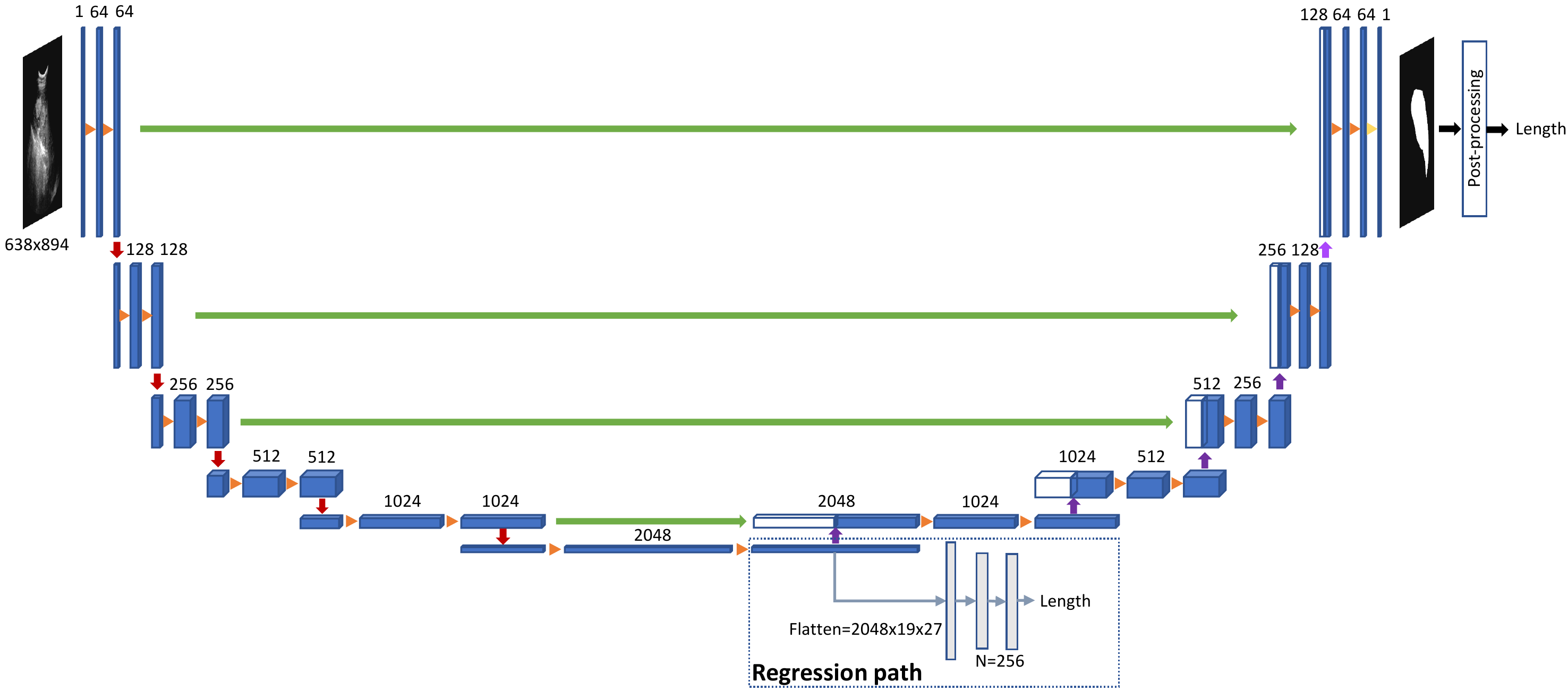}}
	\subfigure[VGG-19 based] {\includegraphics[width=1\textwidth]{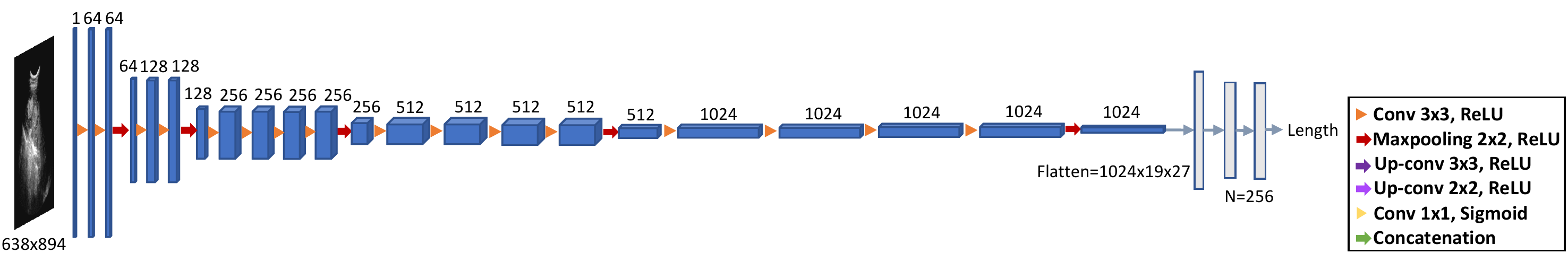}}
    \caption{Diagram showing the architectures for the three proposed models. (a) The first model is based upon a U-Net architecture, which performs a segmentation task. This is followed by CCA/PCA based postprocessing to estimate length. The encoding path of the U-net is also used in the second model, and combined with the extra fully connected layers within the dotted frame, which perform direct regression of spleen length. Note that this model does not include the decoding path of the U-net. (b) In the third model the same fully connected layers as in (a) are added at the end of a VGG-19 architecture to estimate length.} \label{Fig1}
\end{figure*}

Two types of approach were investigated for estimating the length of the spleen from the images, as illustrated in Fig.~\ref{Fig1}. The first type (network outside the dotted frame in Fig.~\ref{Fig1}(a)) was based on an automated deep learning segmentation followed by length estimation (see Section \ref{sect:segapproach}).
In the second type of approach (see Section \ref{sect:regressapproach}) we estimated length directly using deep regression models. Two different techniques for length regression were investigated (dotted frame in Fig.~\ref{Fig1}(a) and Fig.~\ref{Fig1}(b)).

\subsection{Segmentation-based approach}
\label{sect:segapproach}

We used a U-Net based architecture \cite{ronneberger2015u} to automatically segment the spleen. Because of the size of our input images, we added an additional downsampling block to further compress the encodings in the latent space compared to the original U-Net.
We applied dropout with probability 0.5 to the bottle-neck of our network. Each convolutional layer was followed by batch normalisation. We used the ReLU activation function for each convolutional layer and the sigmoid activation function in the output layer to classify each pixel as either spleen or background. A Dice loss function was used to penalise the difference between predicted labels and ground truth labels.

Based on the segmentation output of the U-net, we performed connected components analysis (CCA) to preserve the largest foreground (i.e. spleen) region. We then applied principle components analysis (PCA) on the coordinates of the identified spleen pixels to find the longest axis. The spleen length was then computed as the range of the projections of all spleen pixel coordinates onto this axis.

\subsection{Regression-based approach}
\label{sect:regressapproach}

We investigated two different models for direct estimation of spleen length.

In the first model the encoding part of the U-net was used to estimate a low-dimensional representation of spleen shape. We then flattened the feature maps in the latent space and added two fully connected layers  (dotted frame in Fig. \ref{Fig1}(a)) with number of nodes N $=$ 256. Each fully connected layer was followed by batch normalisation.  Note that this model does not include the decoding path of the U-net.

We compared this model to a standard regression network (Fig.~\ref{Fig1}(b)), the VGG-19~\cite{simonyan2014very}.
To enable a fairer comparison between the two regresssion approaches, the same number of nodes and layers were used for the fully connected layers in the VGG-19 as were used for the first regression network. Batch normalisation was applied after each layer.

The mean square error loss was used for both models. 


\section{Experiments and Results}
\subsection{Materials}
\label{sect:materials}
A total of 108 2D ultrasound images from 93 patients (aged 0 to 18) were used in this study. All patients were children with SCD and received professional clinical consultation prior to ultrasound inspection. Up to 4 ultrasound images were obtained from each patient during a single examination. Ultrasound imaging was carried out on a Philips EPIQ 7. During the process of imaging, an experienced sonographer followed the standard clinical procedure by acquiring an ultrasound plane that visualised the longest axis of the spleen and then manually marked the starting and ending points of the spleen length on each image. To remove the manual annotations on the acquired images, we applied biharmonic function-based image inpainting on all images~\cite{damelin2018surface}. The images were then manually cropped to a 638 $\times$ 894 pixel region of interest covering the entire spleen.

In addition to the manual length measurements made by the original sonographer (E1), manual measurements were made  from the acquired ultrasound images (with annotations removed) by two further experts (E2 and E3) in order to allow quantification of inter-observer variability.
Due to variations in pixel size between the images, all manual spleen length measurements (in millimetres) were converted to pixels for use in training the regression models. 

To train the segmentation-based approach, the spleen was manually segmented in all images by a trained observer using ITK-SNAP~\cite{yushkevich2006user}.

\subsection{Experimental Setup}
\label{sect:Experiments}
Data  augmentation  was  implemented in training all models,  consisting  of  intensity  transformations (adaptive histogram equalisation and gamma correction) and spatial transformations  (0 to 20 degree random rotations). We intensity-normalised all images prior to using them as input to the networks. For evaluation, we used a three-fold nested cross-validation. For each of the three folds of the main cross-validation, the 72 training images were further separated into three folds, and these were used in the nested three-fold cross-validation for hyperparameter optimisation (weight decay values $10^{-6}$, $10^{-7}$ and $10^{-8}$). The model was then trained on all 72 images based on the best hyperparameter and tested on the remaining 36 images in the main cross-validation. This process was repeated for the other two folds of the main cross-validation.

For the second model (i.e. direct estimation using the U-net encoding path), we investigated whether performance could be improved by transferring weights from the U-net trained for the segmentation task. Therefore, in our experiments we compared four different techniques based on the three models outlined in Fig.\ref{Fig1}:
\begin{enumerate}
    \item Segmentation-based approach followed by postprocessing (SB).
    \item Direct estimation based on the encoding path of the U-net, without weight transfer from the U-net trained for segmentation (DE).
    \item The same direct estimation approach with weight transfer from the segmentation U-net (DEW).
    \item The VGG-19 based direct estimation approach (VGG).
\end{enumerate}

\subsection{Implementation Details}
\label{sect:Implementation}
The proposed models were all implemented in PyTorch and trained using an NVIDIA TITAN RTX (24GB). The learning rate was set to $10^{-5}$ for all models. All models were trained using the Adam optimiser, with a batch size of 4 due to memory limitations. 

\subsection{Results}
\label{sect:Results}
The results are presented in Table~\ref{tab1}. This shows the following measures:
\begin{itemize}
    \item PLE (Percentage Length Error): The percentage error in the length estimate compared to the ground truth.
    \item R: The Pearson's correlation between length estimates and the ground truth.
    \item Dice: For SB only, the Dice similarity metric between the ground truth segmentation and the estimated segmentation.
    \item HD (Hausdorff Distance): For SB only, the general Hausdorff distance between estimated and ground truth segmentations.
\end{itemize}

\begin{figure*}[!htb]\center
	\includegraphics[width=1\textwidth]{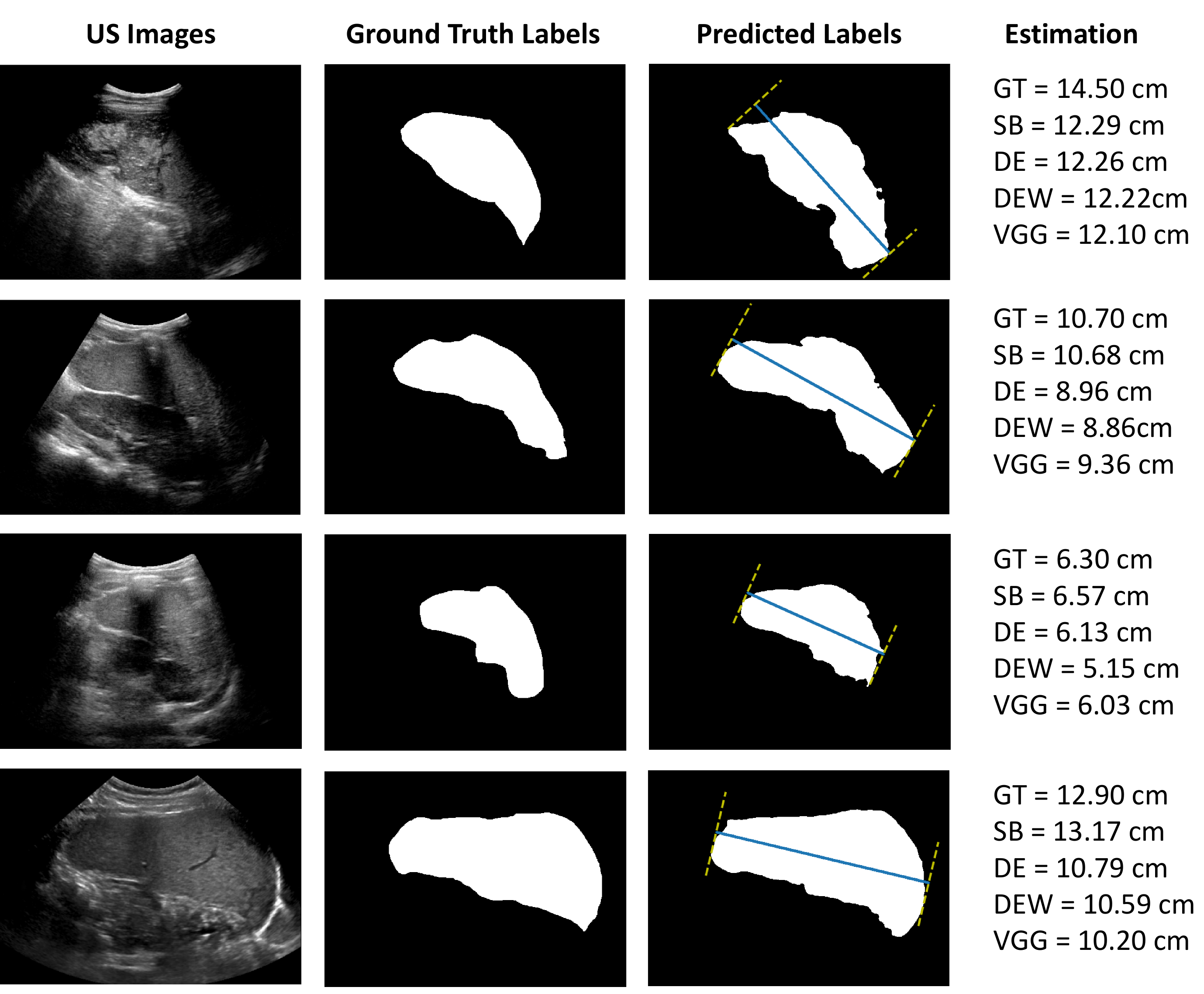}
    \caption{Visualisations of ultrasound images, ground truth labels (i.e. segmentations) and predicted labels after CCA. The blue line segments in the predicted labels indicate the PCA-based length estimates. Ground truth lengths and estimated lengths made using the different approaches are presented in the fourth column. Each row represents a different sample image.} \label{Fig2}
\end{figure*}

For all measures, the ground truth was taken to be the original manual length measurement made by E1 and the measures reported in the table are means over all 108 images. We present some examples of images and segmentations in Fig. \ref{Fig2}, along with ground truth lengths and estimations made by the different models.

Table~\ref{tab1} also shows the agreement between the length measurements made by the three human experts (E1, E2 and E3).

The results show that the segmentation-based approach outperformed the other three approaches and its performance is close to the level of inter-observer variability between the human experts. The direct estimation methods perform less well, and we find no improvement in performance from weight transfer.

\begin{table}[!t]
\caption{Comparison between results for segmentation-based estimation (SB), direct estimation (DE), direct estimation with weight transfer (DEW) and VGG-19 based direct estimation (VGG). PLE: Percentage Length Error, R: Pearson's correlation, Dice: Dice similarity metric, HD: general Hausdorff Distance. All values are means over all 108 images. The final three columns quantify the level of inter-observer variability between the three experts E1, E2, and E3.}\label{tab1}
\begin{tabular}{lccccccc}
\hline
\textbf{Methods} & \textbf{ SB } & \textbf{ DE } & \textbf{ DEW } & \textbf{ VGG } & \textbf{ E1 vs E2 } & \textbf{ E1 vs E3 } & \textbf{ E2 vs E3 } \\ \hline
PLE   & 7.42\%    & 12.80\%    & 12.88\%    & 12.99\%    & 5.47\%     & 5.52\%     & 6.34\%     \\ \hline
R   & 0.93    & 0.86      & 0.87      & 0.88      & 0.94      & 0.97      & 0.93      \\ \hline
Dice    & 0.88           & \textbf{-} & \textbf{-} & \textbf{-} & \textbf{-} & \textbf{-} & \textbf{-} \\ \hline
HD (mm) & 13.27            & \textbf{-} & \textbf{-} & \textbf{-} & \textbf{-} & \textbf{-} & \textbf{-} \\ \hline
\end{tabular}
\end{table}

\section{Discussion and Conclusion}
In this work, we proposed three models for automatic estimation of spleen length from ultrasound images. To the best of our knowledge, this is the first attempt to perform this task in a fully automated way. We first adjusted the U-Net architecture and applied post-processing based on CCA and PCA to the output segmentation to estimate length. We also proposed two regression models, one based on the U-Net encoding path and one based on the well-established VGG-19 network.

Our results showed that the segmentation-based approach (SB) had the lowest PLE and the highest correlation. The performance of this approach was close to the level of human inter-observer variability. The PLE and correlation for the first direct estimation approach with or without weight transfer (DE and DEW) were similar. This indicates that weights learned from the segmentation task do not help to improve the performance of the direct estimation task. This is likely due to a strong difference between the optimal learnt representations for segmentation and length estimation tasks. The results produced by the two regression-based models (DE/DEW and VGG) are also similar. However, compared to the U-Net encoding path direct estimation network (DE, DEW), VGG-19 has a relatively small number of parameters (178,180,545 vs. 344,512,449) but achieves similar results, which demonstrates the potential of the VGG-19 model for this length estimation task.

Although we have achieved promising results in this work, there are some limiting factors that have prevented the results from being even better. The first is that some of our training data suffer from the presence of image artefacts. The second is that the spleen and its adjacent fat have quite low intensity contrast. These two factors combined make measuring the length of the spleen very challenging, and it is likely that human experts use knowledge of the expected location and anatomy of the spleen when making manual measurements to overcome a lack of visibility of the spleen. Finally, we have a limited amount of data (108 images). In the future, we aim to obtain more ultrasound images. With a larger database of training images, we believe the performance of the segmentation-based approach has the potential to reach or surpass that of human experts. We also plan to investigate alternative architectures to exploit possibly synergies between the segmentation and length estimation tasks.

\section*{Acknowledgements}
This work was supported by the Wellcome/EPSRC Centre for Medical Engineering [WT 203148/Z/16/Z]. The support provided by China Scholarship Council during PhD programme of Zhen Yuan in King's College London is acknowledged.

%
%
%

\bibliographystyle{splncs04}

\end{document}